\documentstyle[aps,prb,floats,epsf]{revtex}
\begin{document}

\twocolumn[\hsize\textwidth\columnwidth\hsize\csname
@twocolumnfalse\endcsname

\title{Maximally-localized Wannier functions for disordered systems:
application to amorphous silicon}
\author{Pier Luigi Silvestrelli,$^1$ Nicola Marzari,$^2$
David Vanderbilt,$^2$ Michele Parrinello$^1$}
\address{$^1$Max-Planck-Institut f\"ur Festk\"orperforschung,
Heisenbergstr. 1, 70569 Stuttgart, Germany\\}
\address{$^2$Department of Physics and Astronomy, Rutgers 
University, Piscataway, NJ 08854-8019, USA \\} 

\date{\today}
\maketitle

\begin{abstract}
We use the maximally-localized Wannier function method
to study bonding properties in amorphous silicon. 
This study represents, to our knowledge, 
the first application of the Wannier-function analysis 
to a disordered system. Our results show that, in the presence of
disorder, this method is extremely helpful in providing an unambiguous
picture of the bond distribution. In particular, defect configurations 
can be studied and characterized with a novel degree of accuracy that
was not available before.

\end{abstract}

\vspace{3mm}

\vskip2pc]

\narrowtext

\section{Introduction} 
Since their introduction in 1937 Wannier functions\cite{Wannier} 
have played an important role in the theoretical study
of the properties of periodic solids.
Moreover the representation of the electronic ground state
of periodic systems in terms of localized Wannier or
Wannier-like orbitals has recently attracted 
considerable attention due to the development
of ``order-N'' methods\cite{orderN} and to the formulation
of the modern theory of electronic polarization\cite{polarization}. 
In the case of finite systems, localized orbitals are
widely used to describe and understand chemical concepts 
such as bonds, lone-pair orbitals, and valence-electron charge distributions;
different criteria\cite{criteria,Boys,ELF} have been developed for
producing optimum localized orbitals.

In periodic systems the determination of localized orbitals or
Wannier functions is much less trivial\cite{methods,Marzari}.
Recently, Marzari and Vanderbilt\cite{Marzari} 
developed a very practical method for generating maximally-localized 
Wannier functions starting from the knowledge of the occupied 
Bloch states. This amounts to the generalization for periodic systems of the
Boys' localized-orbital method\cite{Boys} that is commonly used in quantum
chemistry. The new technique has been successfully applied
to crystal systems and small molecules\cite{Marzari}.
Here we apply for the first time the same procedure
to a disordered system, namely amorphous silicon.
We show that, also in this case, the Wannier functions are extremely
useful in providing a clear description of the
relevant electronic and bonding properties.
They also help in eliminating many of the ambiguities that are usually
associated with identifying defects in a disordered system.

\section{Method} 
The Wannier functions\cite{Wannier} are defined in terms of a 
unitary transformation of the occupied Bloch orbitals. Even for the case of
a single band, the Wannier functions are not uniquely defined, due to the 
arbitrary freedom in the phases of the Bloch orbitals. In the 
multiband case this freedom becomes more general, and includes the choice
of arbitrary unitary transformations among all the occupied orbitals at
every point in the Brillouin zone (BZ).
Marzari and Vanderbilt\cite{Marzari} resolve this
indeterminacy by requiring that the total spread of the Wannier functions
\begin{equation}
S = \sum_n \left( \left<r^2\right>_n - \left<{\bf r}\right>^2_n \right)\;,
\label{spread}
\end{equation}
be minimized in real space, in analogy with Boys' criterion\cite{Boys} for
finite systems. 
In Eq.~(\ref {spread}) $\left<...\right>_n$ indicates
the expectation value with respect to the $n$-th Wannier function $w_n({\bf r})$.
Marzari and Vanderbilt\cite{Marzari} have discussed how to 
properly define the operators ${\bf r}$ and $r^2$ in a periodic system and
have detailed the procedure to determine the functions $w_n({\bf r})$ 
for a general $k$-point sampling of the BZ.
Since we have in mind applications to large and 
disordered systems, we restrict
ourselves here to the case of $\Gamma$-point only sampling of the BZ.

Our optimization procedure is closely related to that described
in Appendix A of Ref. \onlinecite{Marzari}.
We report explicitly the formulas to be used in a calculation with a cubic 
supercell of side $L$, which is the case of our simulation.
The minimum spread criterion of Eq.~(\ref {spread})
is equivalent to the problem of maximizing the functional
\begin{equation}
\Omega = \sum_n \left(|X_{nn}|^2+|Y_{nn}|^2+|Z_{nn}|^2\right)\;,
\label{omegamax}
\end{equation}
where $X_{mn}=\left< w_m | e^{-i{2\pi \over L} x} | w_n \right>$. 
Similar definitions for $Y_{mn}$ and $Z_{mn}$ apply.
Maximization of $\Omega$ is performed using a steepest descent (SD) algorithm.
We start the procedure by
constructing the new matrices $X^{(1)}$, $Y^{(1)}$ and $Z^{(1)}$ via
the unitary transformations $X^{(1)}={\rm exp}(-A^{(1)})X^{(0)}
{\rm exp}(A^{(1)})$ (and similarly for $Y^{(1)}$ and $Z^{(1)}$), where 
$X^{(0)}_{mn}=\left< w^{(0)}_m | e^{-i{2\pi \over L} x} | w^{(0)}_n \right>$ and
$w^{(0)}_n({\bf r})=\psi_n({\bf r})$ are the Kohn-Sham (KS) orbitals
obtained after a conventional electronic structure calculation.
$A^{(1)}$ is an antihermitian matrix corresponding to a finite step
in the direction of the gradient of $\Omega$ with respect to all the
possible unitary transformations given by ${\rm exp}(-A)$:
$A^{(1)}=\Delta t \,(d\Omega/dA)^{(0)}$, where $\Delta t$ is the
conventional SD ``time-step''. The gradient $d\Omega/dA_{mn}$ is given by
the sum of $[X_{nm}(X_{nn}^*-X_{mm}^*)-X_{mn}^*(X_{mm}-X_{nn})]$
and the equivalent terms with $Y$ and $Z$ substituted in place of $X$. 
The process is repeated for many SD iterations up to convergence in 
the $\Omega$ functional.
The maximally-localized Wannier functions are then
given by $w_n({\bf r})=\Pi_i \, {\rm exp}(-A^{(i)})\psi_n({\bf r})$.
The coordinate $x_n$ of the $n$-th Wannier-function
center (WFC) is computed using the formula
\begin{equation}
x_n = -{L\over {2\pi}}{\rm Im}\; {\rm ln} 
\left< w_n | e^{-i{2\pi \over L} x} | w_n \right>\;,
\label{rcenter}
\end{equation}
with similar definitions for $y_n$ and $z_n$.
Eq.~(\ref {rcenter}) has been shown by Resta\cite{Resta} 
to be the correct definition of the expectation value of 
the position operator for a system with periodic boundary conditions.
The computational effort required in Eqs.~(\ref {omegamax}) and
(\ref {rcenter}) is negligible, once the scalar products needed to
construct
the initial $X^{(0)}$, $Y^{(0)}$ and $Z^{(0)}$ have been calculated.

\section{Results and discussion} 
Amorphous silicon
has been one of the first systems studied\cite{Stich} with the 
Car-Parrinello molecular dynamics method\cite{CP}.
We analyse here some selected configurations taken from a molecular-dynamics
simulation performed by Chiarotti\cite{Chiarotti}.
These configurations have been obtained by quenching from 
the liquid state a sample of 64 Si atoms, contained in a cubic
supercell of side 10.86 \AA\ and periodically repeated in space.
The total length of the simulation run was about 10 ps.
Since our study is based on the local density approximation (LDA)
to density-functional theory (DFT), each orbital is occupied 
twice and unpaired spin defects cannot be observed.
However, such defects are expected to have a low density, as suggested
by electron spin resonance experiments\cite{ESR_Si}.

In amorphous silicon most of the atoms are tetrahedrally bonded
($sp^3$ hybridized); however different kinds of defects can be
present and have been proposed in the literature
\cite{Stich,Anderson,Adler,Elliott,Fedders}:
twofold coordinated atoms forming spinless, neutral defects;
threefold coordinated atoms with neutral or 
charged dangling bonds; fourfold
coordinated atoms characterized by stretched (``weak'') bonds
and by bond angles that are rather far from the tetrahedral angles;
and fivefold coordinated atoms (``floating bonds'').

The usual analysis of the bonding properties is based on the coordination
number, i.e. the number of atoms lying inside a 
sphere of a chosen radius $r_c$ centered on the selected atom.
Such a simplified structural
analysis is sensitive to the value chosen for $r_c$. More 
importantly, it is also completely blind to the electronic charge distribution,
which ought to be important to any description of chemical bonding.
Analysis of the full charge distribution and bonding in terms of the
Wannier functions is rather complex. However, as we show below, the
knowledge of the positions of the WFCs suffices to capture most of 
the chemistry of the system and to identify its defects.

\begin{figure}
\epsfxsize=3.2 truein
\centerline{\epsfbox{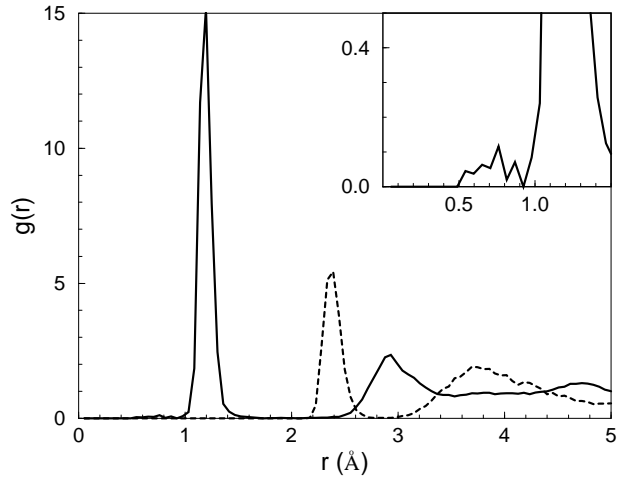}}
\vskip 0.2truein
\caption{Si-Si (dashed line) and Si-WFC (solid line) pair correlation 
functions. The detailed structure, in the range 0.0--1.5 \AA, is shown
in the inset. The data have been obtained by 
averaging over 17 configurations of the MD simulation.}
\label{gr-Si}
\end{figure}
The ionic structure of a disordered system is best described in terms of
correlation functions. We treat here the centers of the localized Wannier functions
as a second species of classical particles, and 
we regard amorphous silicon as a statistical assembly of these two kinds of 
particles, the Si ions and the WFCs.
In Fig.\ \ref{gr-Si} we show $g(r)$, the standard Si-Si pair correlation 
function, together with $g_w(r)$, the Si-WFC pair correlation function. 
As can be seen, $g(r)$ and $g_w(r)$ exhibit strong peaks at $\simeq 2.4$
and at $\simeq 1.2$ \AA, respectively, thus indicating that the 
electronic charge
is mostly localized in the middle of the Si-Si bonds, as expected in 
a covalent-bonded system. 
The spread in the peaks indicates the disorder-induced strain in the
Si-Si bond.
However, $g_w(r)$ shows some structure also
for values of $r$ in the range 0.5--1.0 \AA $\,$ (see inset). 
This means that a few WFCs
are anomalous in being very close to the Si ions.
This behaviour represents a clear indication of the presence of defects
in the system.

If the coordination number is computed by integration of $g(r)$ up to
$r_c=2.80$ \AA\ (the position of the first minimum), we find that, on average,
96.5 \% of the Si ions are fourfold coordinated, while 3.5 \% are 
fivefold coordinated. Therefore, according to an analysis
restricted to the ionic coordinates alone,
all defects in the system are identified as
fivefold-coordinated Si ions. This is 
in agreement with the findings of previous 
ab-initio simulations\cite{Stich}.
However, when we choose a coordination criterion based on the Wannier function 
representation, we get rather different results.
We will say that a bond exists between two Si ions when they share
a common WFC located within $r_w=1.75$ \AA\ of each ion,
$r_w$ being the position of the first minimum of $g_w(r)$. With
this convention,
we find that 97.5 \% of the Si ions are fourfold bonded; of the remaining
ions, only $\sim 0.6$ \% have 5 bonds, while the others are more or less equally 
subdivided into twofold-bonded and threefold-bonded ions.
Therefore, although the total density of defective atoms that we obtain
is similar to that coming from the bare coordination analysis, 
the nature of the defects appears to be different.

\begin{figure}[!ht]
\epsfxsize=3.2 truein
\centerline{\epsfbox{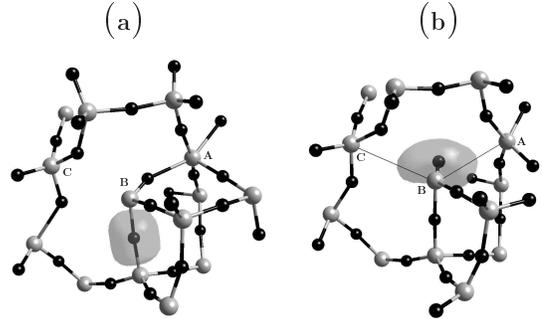}}
\vskip 0.2truein
\caption{Snapshots of 2 different configurations of the MD simulation
of amorphous silicon. Large grey balls denote Si ions, while small black
balls denote WFCs. A, B and C label the Si ions, whose bonding properties
are discussed in the text. For clarity only the Si ions and WFCs lying
within 4 \AA $\,$ of ion B are shown. 
We also plot the isosurface densities
$\rho_n({\bf r})=|w_n({\bf r})|^2$ corresponding to
a normal covalent Wannier function in (a), and to an anomalous
Wannier function close to ion B in (b).
Thin lines in (b) indicate directions of possible bonds of ion B with ions A 
and C.}
\label{snap-Si}
\end{figure}

This fact is best illustrated by looking at the bonds formed among a 
few ions, in selected configurations of the molecular-dynamics
simulation.
In Fig.\ \ref{snap-Si}(a) ion A is fivefold coordinated and 
has 5 bonds, while ion B is fourfold coordinated but has
3 bonds only. In fact no WFC is found between ion B and ion C.
Notice that the bond
between ion A and ion B is somehow anomalous since the distance from the
corresponding WFC to ion B 
is considerably smaller than to ion A (0.87 and 1.56 \AA\ respectively), and the 
A--B bond appears to be distorted. 
As the ions move, the electronic configuration also changes, and in fact
we find that after about 10 ps the WFC located between ion A
and ion B comes still closer to ion B 
(see Fig.\ \ref{snap-Si}(b)).
The distance is reduced to 0.57 \AA,
in such a way that the A--B bond is broken or at least severely weakened.
In this configuration, according to our criterion, ion A is fourfold 
bonded, while ion B has only 2 bonds. Interestingly, the
twofold-bonded atom was proposed by Adler\cite{Adler} as the
lowest-energy defect in amorphous silicon.
The defect we observe in Fig.\ \ref{snap-Si}(b) probably represents a 
transient state in which ion B breaks the bond with ion A and tries
to form a new bond with a different nearest-neighbour, possibly ion C.
This conclusion is supported by the fact that the direction of the vector 
connecting ion B to the anomalous WFC is intermediate between the
B--A and B--C directions. Further confirmation comes from 
inspection of the isosurface of the electronic charge distribution
associated with the anomalous Wannier function. Notice that the density
profile of this Wannier function is different from the 
one associated with a normal covalent bond, as
shown in Fig.\ \ref{snap-Si}(a).
To determine whether a new bond with ion C is really formed or
whether the configuration of ion B in Fig.\ \ref{snap-Si}(b) is stable
would require a longer simulation, which is beyond the scope of the
present work.

We stress again that the interesting transformation in the bonding properties
of ions A and B, which we have described using the Wannier function analysis,
cannot be detected using the simple coordination number criterion.
In fact, in the configuration of Fig.\ \ref{snap-Si}(b), the coordination 
numbers of ion A and B are the same as in the configuration of 
Fig.\ \ref{snap-Si}(a), although the A--B distance is significantly increased
from 2.38 to 2.59 \AA. 

The Wannier function analysis allows a description of the electronic charge
distribution in terms of well-defined, localized 
functions; the clear 
representation of the bonding properties of the system based on the
positions in real space of the WFCs can then be followed
by a more quantitative study.
For instance, one can relate specific features of the electronic
density of states to a particular Wannier function $w_n({\bf r})$, 
defining a ``projected density of states''

\begin{equation}
N_n(E)=\sum_m |\left< w_n |\psi_m \right>|^2 \delta(E-E_m)\;,
\label{projection}
\end{equation}
where $\psi_m$ and $E_m$ are the KS eigenvectors and eigenvalues.
\begin{figure}
\epsfxsize=3.2 truein
\centerline{\epsfbox{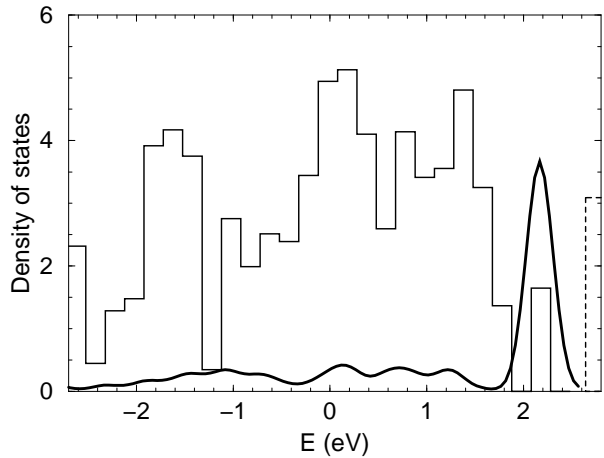}}
\vskip 0.2truein
\caption{Histogram of the electronic density of states (thin line, the
dashed portion indicating the conduction band)
compared with the projected density of states $N_n(E)$ (thick line) 
corresponding to the anomalous WFC of Fig. 2(b) (see text for definition). 
The histogram has relatively large fluctuations because only a single ionic
configuration is considered. The curve $N_n(E)$ has been smoothed with a
gaussian broadening.}
\label{partedos}
\end{figure}
As can be seen from Fig.\ \ref{partedos}, the $N_n(E)$ function
associated to the anomalous WFC of Fig.\ \ref{snap-Si}(b) exhibits a 
strong peak which matches the peak in the electronic density
of states located above the valence-band edge. This peak can be associated
with the highest-energy occupied KS state. We have therefore a clear
example in which the electronic states of the structural defects present 
in our sample are introduced into the energy gap.

Another important advantage of the use of the Wannier functions
is that a precise calculation of the localization degree of the
electronic charge is possible, in contrast with previous 
approximate estimates\cite{Stich,Fedders}.
In fact one can easily compute the quantity 

\begin{equation}
\sigma_n = \sqrt{\left<r^2\right>_n - \left<{\bf r}\right>^2_n}\;,
\label{sigma}
\end{equation}
which corresponds to the spread in real space of the Wannier function
$w_n({\bf r})$. Considering again the anomalous WFC of 
Fig.\ \ref{snap-Si}(b), we find that $\sigma_n = 1.94$ \AA, a value
significantly larger than the average spread obtained by considering all 
the WFCs, $\bar\sigma= 1.38$ \AA.
We find therefore that the Wannier function
associated to the defect state is less localized that the Wannier functions
associated to normal states. 
This interesting result confirms, in a quantitative way, the qualitative
observation of Fedders, Drabold and Klemm\cite{Fedders} who pointed out
that the defect states can be far less localized than expected. 
 
\section{Conclusions} 
In conclusion, we have described a novel application of the
Wannier function analysis to a disordered system.
We have shown that a simple geometrical analysis of the positions
of the WFCs is already sufficient to extract useful information
about the bonding properties of the ions, particularly in those
interesting defective configurations which, using traditional approaches,
can only be studied in a very crude way.
In addition, since in our method the localized Wannier functions
are explicitly available, a quantitative analysis, 
which allows us to estimate accurately the 
degree of localization of the electronic charge, is possible. 
It also allows us to clarify the nature
of the relevant features in the electronic density of states.

An additional advantage of the current approach of analyzing the
dynamics of the system of ions plus WFCs is that it should be
possible to extract information about the dipolar fluctuations,
i.e. about the dielectric response function $\chi(\bf{k},\omega)$.  In fact
the present approach is closely related to that of Ref. \onlinecite{IRprr}
where, however, only the $k=0$
response relevant to optical probes is computed, so that
only the macroscopic polarization is needed at each time step.
The current approach opens the possibility of extracting some
local (i.e. $k$-dependent) information as well.

\acknowledgments
We thank M. Boero and M. Bernasconi for useful discussions.

\vfill
\eject

\end{document}